\documentclass[12pt]{iopart}

\usepackage{graphicx}% Include figure files
\usepackage{dcolumn}% Align table columns on decimal point
\usepackage{bm}% bold math
\usepackage{hyperref}% add hypertext capabilities
\usepackage{graphicx}
\usepackage{physics}
\usepackage{xargs}  
\usepackage[pdftex,dvipsnames]{xcolor}
\usepackage{hyperref}
\usepackage{eqparbox}
\usepackage{tikz}
\usetikzlibrary{quantikz}
\usepackage{multirow}
\usepackage{tabularx}

\usepackage{cite}
\newcommand\numberthis{\addtocounter{equation}{1}\tag{\theequation}}

\newcommand{\cosphi}{\ensuremath{\cos\pqty{\phi}}}
\newcommand{\sinphi}{\ensuremath{\sin\pqty{\phi}}}
\newcommand{\ficcmax}{\ensuremath{FI_{CC}^{\max}}}
\newcommand{\fiamax}{\ensuremath{FI_A^{\max}}}
\newcommand{\fibmax}{\ensuremath{FI_B^{\max}}}
\newcommand{\meqbox}[2]{\eqmakebox[#1][r]{$\displaystyle#2$}}

\begin{document}

\title{Lossy SU(1,1) interferometers in the single-photon-pair regime}

\author{Matteo Santandrea,$^{1}$ $^{\ast}$ Kai-Hong Luo, $^{1}$ $^{\dagger}$ Michael Stefszky,$^{1}$ Jan Sperling,$^{2}$ Harald Herrmann,$^{1}$ Benjamin Brecht,$^{1}$ and Christine Silberhorn$^{1}$}

\address{$^{1}$Paderborn University, Department of Physics,
Integrated Quantum Optics, Institute for Photonic Quantum Systems (PhoQS), Warburgerstr. 100, D-33098 Paderborn, Germany}

\address{$^{2}$
Paderborn University, Department of Physics,
Theoretical Quantum Science, Institute for Photonic Quantum Systems (PhoQS), Warburgerstr. 100, D-33098 Paderborn, Germany
}

 \eads{\mailto{$^{\ast}$matteo.santandrea@upb.de}, \mailto{$^{\dagger}$khluo@mail.uni-paderborn.de}}

%% email address is required

\date{\today}


\begin{abstract}
The success of quantum technologies is intimately connected to the possibility of using them in real-world applications. 
To this aim, we study the sensing capabilities of quantum SU(1,1) interferometers in the single-photon-pair regime and in the presence of losses, a situation highly relevant to practical realistic measurements of extremely photosensitive materials. 
We show that coincidence measurement can be exploited to partially mitigate the effect of losses inside the interferometer.
Finally, we find that quantum SU(1,1) interferometers are capable of outperforming classical SU(2) systems when analogous real-world conditions are considered.
\end{abstract}

%\maketitle
%%%%%%%%%%%%%%%%%%%%%%%%%%  body  %%%%%%%%%%%%%%%%%%%%%%%%%%

\section{Introduction}
Precise determination of physical quantities, such as dispersion and absorption, is often achieved via accurate phase measurement in an interferometric setup. In the never-ending quest to improve the maximum phase resolution achievable, the new class of SU(1,1) interferometers was introduced in 1986 by Yurke \textit{et al}. \cite{Yurke1986}.
These systems can be described as Mach-Zehnder interferometers (MZIs) whose input/output beam splitters have been replaced by two nonlinear elements.

This class of interferometers has been widely studied in the context of quantum metrology \cite{Ou2020a}, particularly in the high-gain regime, i.e., when a high mean photon number, $\expval{n}\gg 1$, is generated inside the interferometer.
In this operation regime, the achievable precision of SU(1,1) interferometers can scale as $\sim 1/\expval{n}$, the so-called Heisenberg scaling. 
Hence, these systems can outperform linear MZI with classical input light, whose scaling is limited to the standard quantum limit (SQL) $\sim 1/\sqrt{\expval{n}}$.
These interferometers have been theoretically investigated extensively, in particular within the context of Gaussian-state interferometry \cite{Sparaciari2016}, both in the ideal, lossless case \cite{Sparaciari2015a} and in the presence of losses \cite{Ou2012, Marino2012, Giese2017, Hu2018}. 
It was found that they can be particularly robust against detection losses, while losses within the interferometer can have a detrimental impact and even prevent reaching the Heisenberg scaling \cite{Marino2012}.  
Improvement beyond the SQL has been shown experimentally, demonstrating the validity of the theoretical studies so far \cite{Manceau2017a, Du2018}.
Moreover, new variants of Yurke's SU(1,1) interferometer have been proposed and analysed, with particular attention on the properties of these systems when different types of input seeding and detection schemes are considered \cite{Plick2010, Li2016, Anderson2017,Adhikari2018}. 
All this work is currently stimulating an interesting debate over what is, ultimately, the actual resource that allows better-than-SQL resolution in an SU(1,1) interferometer\cite{Gong2017, Caves2020}.

On the other side of the spectrum, these systems have also attracted much attention in the low-gain regime, where $\expval{n} \ll 1$. 
In these conditions, SU(1,1) interferometers do not show Heisenberg scaling \cite{Sparaciari2016, Florez2018}. 
However, they have been utilised to investigate fundamental quantum mechanical concepts, such as state superposition, the quantum Zeno effect, and induced coherence without interaction \cite{Zou1991, Klyshko1993, Herzog1994, Luis1996, Jha2008}. 
Moreover, the correlations of the states generated within an SU(1,1) interferometer have enabled microscopy and spectroscopy with undetected photons \cite{Lemos2014, Kalashnikov2016, Paterova2017, Paterova2018, Vanselow2020, Kviatkovsky2020,  Lindner2020a}, as well as the generation of tailored biphoton entangled states \cite{Riazi2019, Su2019}.
Realisation of integrated SU(1,1) interferometers for metrological application has also been explored recently, both theoretically and experimentally \cite{Ono2019, Ferreri2021}, showing promising results.
Finally, it has recently been demonstrated that it is even possible to achieve phase sensitivity beyond the one provided by SU(1,1) interferometers by cascading several nonlinear elements, in analogy with classical multi-source interference \cite{Paterova2020a}.

Despite the great interest in this system, little work has been done so far to analyse the phase sensitivity of lossy SU(1,1) interferometers and their relation with the SQL, especially when considering non-Gaussian detection schemes.
The most relevant study in this area so far is the work by Michael \textit{et al.} \cite{Michael2021}, whose analysis focuses mainly on the performance of SU(1,1) interferometer with and without a seeded input.
Understanding under which conditions these systems can improve upon the SQL will pave the way to quantum metrology in photon-starved scenarios and enable real-world applications such as quantum-enhanced spectroscopy and microscopy in photosensitive materials.

In this work we study the properties of lossy SU(1,1) interferometers in the low-gain regime.
We analyse the dependence of the classical Fisher information ($FI$) on the losses present in the system and identify the operation conditions under which quantum SU(1,1) interferometers outperform their respective classical SU(2) counterparts. 
Importantly, these regimes of operations coincide with realistic experimental parameters, which highlights the practical advantage offered by SU(1,1) interferometers.

This paper is structured in two main parts. 
In Section \ref{sec:analy_theory}, SU(1,1) interferometers in the low-gain regime are discussed. 
Analytic expressions for the $FI$ of different detection schemes are derived and examined. 
In Section \ref{sec:comparison}, we investigate in which experimental conditions the SU(1,1) interferometer can provide an advantage, when compared to classical MZI.
The results of this comparison show that lossy, low-gain SU(1,1) interferometers can beat the SQL if the internal average transmission is above 50\%.

\section{SU(1,1) in the low-gain regime, click detection, and classical Fisher information.}
\label{sec:analy_theory}
\begin{figure}[tbp]
    \centering
    \vspace{0.4cm}
    \includegraphics[width=0.9\columnwidth]{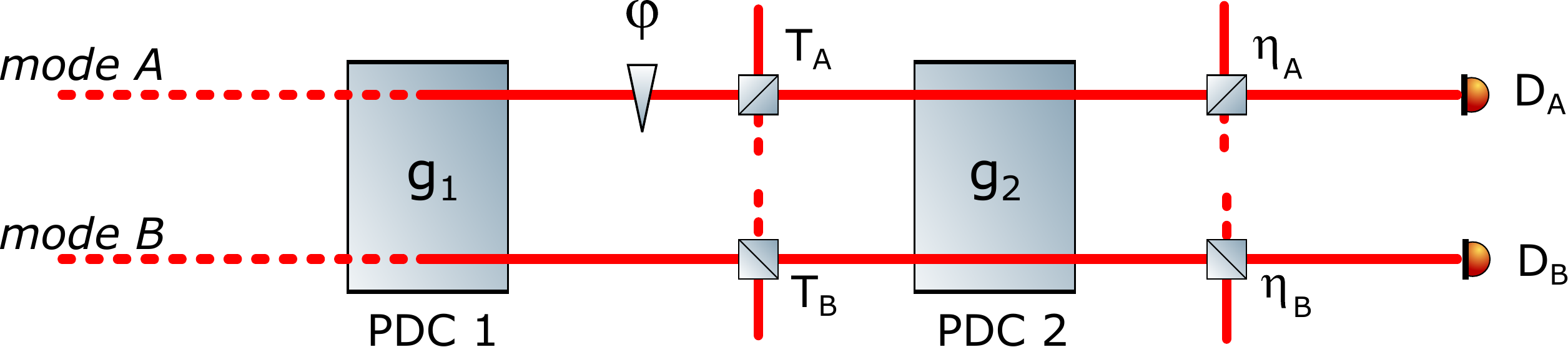}
    \caption{Schematics for the SU(1,1) interferometer with losses and non-ideal detection efficiency.}
    \label{fig:su11_sketch}
\end{figure}

The system under investigation is sketched in Figure \ref{fig:su11_sketch}. The interferometer consists of two cascaded stages that generate photon pairs in two different modes, A and B, via a nonlinear process, e.g., via type-II parametric down-conversion. 
The phase between the two stages is assumed to be constant and equal to zero. 
A phase-shifting element, or sample, is placed between the two stages and imparts a phase shift $\phi$ to mode A.
After the second stage, the generated photons are detected at detectors $D_A$ and $D_B$.
Internal losses $\alpha_{A/B}$ inside the interferometer are modelled with beam splitters placed in the path of modes A and B, which transmit mode A/B with probability $T_{A/B} = 1-\alpha_{A/B}$. 
Without loss of generality \footnote{Placing the internal losses before or after the phase element imparts only constant phase shift to the interference pattern.}, internal losses in mode A are assumed to occur after the phase shifter.
Imperfect detection efficiencies in the output modes --- which also include the external losses after the interferometer --- are similarly modelled with a pair of beam splitters with transmission $\eta_{A/B}$ for mode A/B. 

Throughout most of this work, the interferometer is operated in the low-gain regime, i.e., with $g_{1,2} \ll 1$. 
This means that the mean number of photon pairs generated in each stage is $\expval{n}=\sinh^2{g}\approx g^2\ll 1$.
Moreover, we consider the detectors $D_A$ and $D_B$ to be binary detectors, which cannot discriminate whether more than one photon have impinged on them. 
Note that these type of detectors are also routinely used and readily available, making this restriction meaningful in terms of applicability.
This allows us to derive simple equations describing the system in the single-photon-pair regime. 
To understand the limits of validity of the model, we compare the analytic results with numerical simulations, where we include gains up to $g\sim 1$ to understand the impact of multiphoton components. 
Finally, throughout the paper, we assume $g_1$ to be fixed, i.e. we set the number of photons that, on average, interact with the sample.

Under the assumptions of $g_{1/2}\ll 1$ and click detectors, the click probability for a single detector event $p_{A/B}$ (here onward referred to as singles) and for a coincidence event $p_{CC}$ can be approximated as (see Suppl. Mat.)
\begin{subequations}
\begin{align}
    p_A &= \meqbox{A}{\eta_A\bqty{T_A g_1^2 + g_2^2 + 2 \sqrt{T_A  T_B}g_1 g_2\cosphi},}\numberthis \label{subeq:pa}\\
    p_B &= \meqbox{A}{\eta_B\bqty{T_B g_1^2 + g_2^2 + 2 \sqrt{T_A  T_B}g_1 g_2\cosphi},}\numberthis\label{subeq:pb}\\
    p_{CC} &= \eta_A\eta_B\left[T_A T_B g_1^2 + g_2^2 \right.\notag\\
    & \meqbox{A}{\left.+2 \sqrt{T_A  T_B}g_1 g_2\cosphi\right]}\label{subeq:pcc}.
\end{align}
\label{eq:click_prob}
\end{subequations}

A close inspection of Eq.~\eqref{eq:click_prob} reveals that the click probabilities are determined by the interplay of three different terms.
The first two terms, proportional to $g_1^2$ and $g_2^2$, describe the probability of measuring a photon (or a photon pair, in the case of the coincidences) generated in either the first or the second stage, respectively.
They represent incoherent terms that are always present at the output of the interferometer.
The third term, proportional to $\sqrt{T_AT_B}g_1g_2\cos(\phi)$, describes the interference between the probability of measuring a pair generated in either the first or in the second stage; therefore, this last term is present only when coherence between the two stages is ensured.
Interestingly, this third term is the one that allows phase sensing in this interferometer since it is the only one that depends on the internal phase shift $\phi$.

The ratio between the coherent and the incoherent contributions to the click probabilities is related to how well the two stages interfere and thus can provide interesting information about the interferometer properties and sensing capabilities.
This ratio is the so-called visibility $V$ of the interference pattern, which is formally defined as 
\begin{equation}
    V = \frac{p^{\max}-p^{\min}}{p^{\max}+p^{\min}}.
    \label{eq:visibility_definition}
\end{equation}
From Eq.~\eqref{eq:click_prob}, the visibility of the singles ($V_{A/B}$) and of the coincidences ($V_{CC}$) are given by

\begin{subequations}
\begin{align}
    V_{A} &=  \frac{2\sqrt{T_AT_B}g_1g_2}{T_Ag_1^2 +g_2^2},\label{subeq:visa}\\
    V_{B} &=  \frac{2\sqrt{T_AT_B}g_1g_2}{T_Bg_1^2 +g_2^2},\label{subeq:visb}\\
    V_{CC} &= \frac{2\sqrt{T_AT_B}g_1g_2}{T_AT_Bg_1^2 +g_2^2}.\label{subeq:viscc}
\end{align}
\label{eq:visibilities}
\end{subequations}

The analysis of Eq.~\eqref{eq:visibilities} reveals that the visibilities of the singles is always limited by the internal transmission of the interferometer, i.e., $V_A\leq \sqrt{T_B}$ and $V_B\leq \sqrt{T_A}$.
This means that it is impossible to achieve perfect interference for the singles in a lossy SU(1,1) interferometer.
On the other hand, regardless of the detection efficiency and internal losses, the visibility of the coincidences is always maximized ($V_{CC}=1$) at $g_2^2 = g_1^2T_AT_B$.
Given the importance of this operating point, we refer to the condition $g_1^2 T_A T_B = g_2^2$ as \textit{loss-balanced gains} and an interferometer operated at such a point shall be called a \textit{loss-balanced interferometer}. 

Let us dwell for a moment on the meaning of the condition just derived: in a loss-balanced interferometer, one can always set the internal phase $\phi$ such that the total number of coincidences exiting the interferometer is zero.
In this case, detection losses do not matter and the visibility of the coincidences is thus 100\%.
Such an unexpected behaviour arises thanks to the coherence properties of the SU(1,1) interferometer and suggests that photon-number correlations between the modes of the interferometer can contain more information than a single arm.

A more accurate quantification of the information content in the singles and the coincidence measurement is provided by the classical $FI$. 
For a given state and measurement scheme, one can calculate the classical $FI$ of the measurement, given the probability $p_i$ of measuring $i$ photons, as
\begin{equation}
    FI = \sum_{i} \pqty{\partial_\phi \log{p_i}}^2 p_i.\label{eq:general_fisher_info}
\end{equation}
Since we are considering a scenario where multiphoton components are negligible, one can only measure either a click or no click, corresponding perfectly to the presence of a single photon or no photons, respectively. 
This means that Eq. \eqref{eq:general_fisher_info} can be simplified considering only the two complementary probabilities where the detector clicks ($p$) or does not click ($1-p$).
This simplifies the expression of the $FI$ to
\begin{equation}
    FI = \frac{\pqty{\partial_\phi p}^2}{p\pqty{1-p}} \approx \frac{\pqty{\partial_\phi p}^2}{p},\label{eq:fisher_info_general_formula}
\end{equation}
where the last approximation is valid when $p\rightarrow 0$, as is true in the case considered here.

From Eqs.~\eqref{eq:click_prob} and \eqref{eq:fisher_info_general_formula}, one can calculate the maximum classical $FI$ for singles ($FI^{\max}_{A/B}$) and coincidence measurements ($\ficcmax$)  in a lossy SU(1,1) interferometer,

\begin{subequations}
\begin{align}
    \fiamax &= 2\eta_{A} \bqty{T_Ag_1^2 + g_2^2 -\sqrt{ \pqty{T_Ag_1^2  + g_2^2 }^2 -4 T_A T_B g_1^2g_2^2} },\label{subeq:FAmax}\\
    \fibmax &= 2\eta_{B} \bqty{T_Bg_1^2 + g_2^2 -\sqrt{ \pqty{T_Bg_1^2  + g_2^2 }^2 -4 T_A T_B g_1^2g_2^2} },\label{subeq:FBmax}\\
    \ficcmax & = 2\eta_A\eta_B\bqty{T_AT_Bg_1^2 + g_2^2 -\sqrt{ \pqty{T_AT_Bg_1^2  + g_2^2 }^2 -4 T_A T_B g_1^2g_2^2} }\label{subeq:FCCmax}\\
    & = 2\eta_A\eta_B\bqty{T_AT_Bg_1^2 + g_2^2 -\abs{ T_AT_Bg_1^2  - g_2^2} }\notag\\
     &= \begin{cases}
      4\eta_A\eta_B g_2^2 & \text{if $\frac{g_2}{g_1}< \sqrt{T_A T_B}$},\\
      4\eta_A\eta_B T_A T_Bg_1^2  & \text{if $\frac{g_2}{g_1}\geq \sqrt{T_A T_B}$}.
      \end{cases}\notag
\end{align}
\label{eq:fisher_info}
\end{subequations}

The first two expressions describe the information of modes A and B locally while the latter $FI$ can be thought of as the cross-information between the two modes.

In the following subsections, we analyse these equations to discuss the properties of low-gain lossy SU(1,1) interferometers.

\subsection{Perfect detection efficiency}
\label{subsec:perfect_det_eff}
The expressions in Eq.~\eqref{eq:fisher_info} involve quite a complex interplay between the gains of the two stages, the internal transmissions, and the detection efficiencies. 
To isolate the impact of the internal transmissions on the $FI$, we begin by considering the case of ideal detection, i.e., $\eta_A = \eta_B = 1$.

\begin{figure}[tbp]
    \centering
    \includegraphics[width =\columnwidth]{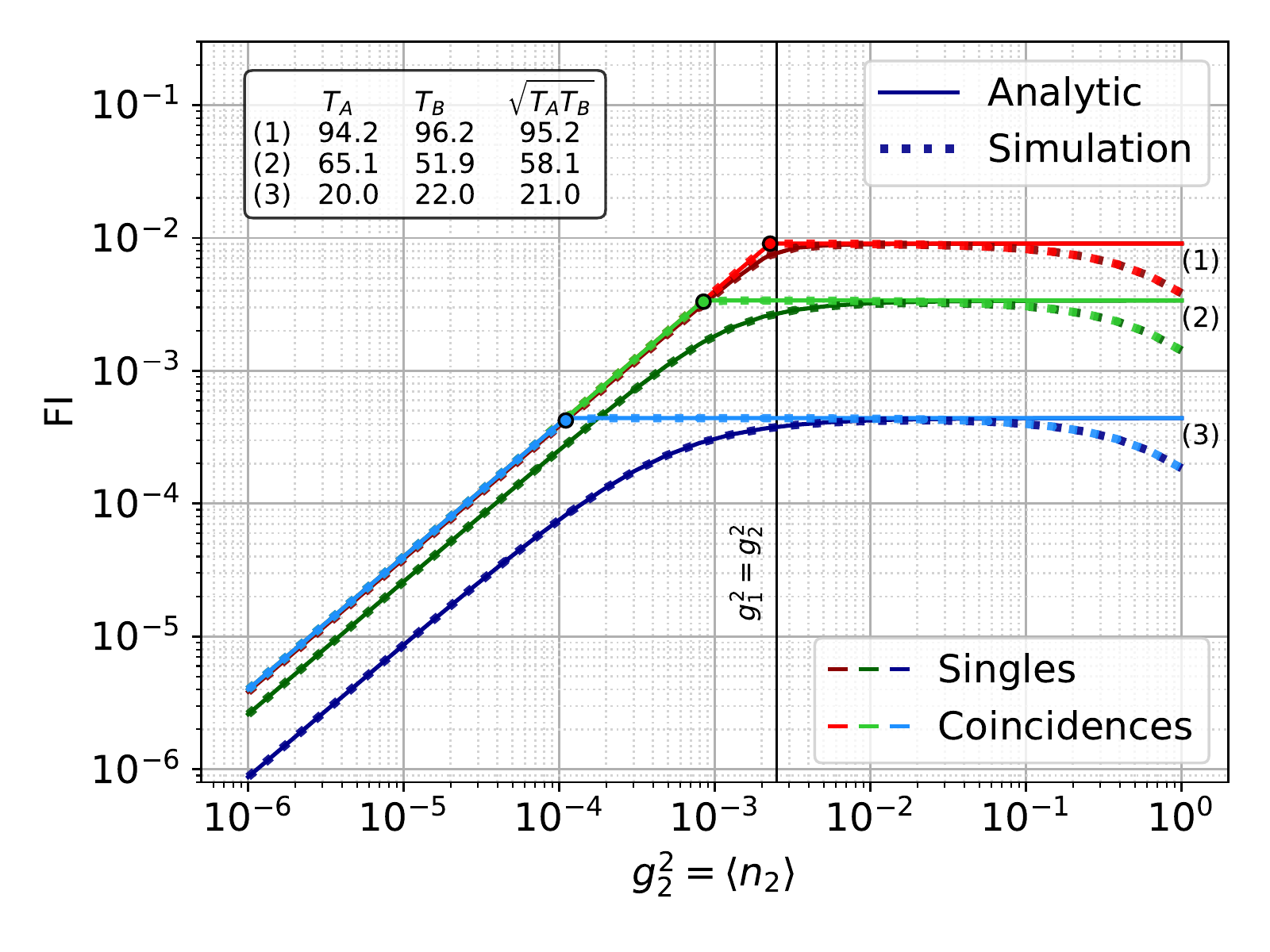}
    \caption{$FI$ of an SU(1,1) interferometer with perfect detection efficiency, gain of the first stage equal to $g_1=0.05$ and internal transmissions equal to $T_A$ and $T_B$ (see the inset in the figure), as a function of the gain $g_2$ of the second stage.
    The solid lines correspond to the analytic theory of Eq. \eqref{eq:fisher_info} while the dashed lines correspond to the numerical simulations (more details in the main text). 
    Lighter lines correspond to the $FI$ of the coincidence measurement, while darker lines correspond to the highest $FI$ of the single measurements. 
    The filled markers correspond to the loss-balanced interferometer, i.e., when $g_2^2 = g_1^2 T_AT_B$, where $V_{CC}=1$.
    The vertical solid black line corresponds to the case where the second stage is pumped with the same gain as the first stage.
    The deviations of the simulations from the analytic theory for $g_2^2>0.3$ are due to the non-negligible contributions of multiphoton effects.}
    \label{fig:internal_losses}
\end{figure}

The behaviour of the $FI$ for different values of internal transmission $T_A$ and $T_B$ is shown by the solid lines in Figure \ref{fig:internal_losses}, where the different colours --- red, green and blue --- correspond to high, medium and low average internal transmission, respectively.
The darker lines correspond to the highest $FI$ of the singles, i.e., $\max\{\fiamax, \fibmax\}$, while the lighter lines correspond to $\ficcmax$.
The analysis of the figure reveals two interesting features. 

The first feature of interest is that the $FI$ of the coincidences is always higher than the $FI$ of the singles, i.e., $F_{CC}^{\max}\geq F_{A/B}^{\max}$. 
Indeed, with some lengthy calculations, it can be proved that this observation is general and independent on the gains and the internal transmissions.
This finding is quite remarkable: the coincidence measurement between the two output modes of the SU(1,1) interferometer provides more information than the one contained in only one of them, as was suggested by the analysis of the visibilities.
This is somewhat unexpected since coincidences are affected by the losses in both arms while the singles only by those in their arm.
In other words, the useful resources in an SU(1,1) interferometer are the number of pairs generated in the first stage.
This finding, along with the observation that one can always obtain perfect visibility for the coincidences, highlights the fact that the information in our SU(1,1) interferometer is not encoded in the photons that pass through the sample in mode A, but in the photon pairs generated in the first stage.
In fact, if we destroy all photon pair correlations generated in the first stage by blocking mode B in the interferometer --- i.e., by setting $T_B = 0$ in Eq. \eqref{eq:click_prob} --- there will still be photons passing through the sample, but we are not be able to observe any interference pattern at the output of the interferometer.

To further clarify why the coincidences always have the highest $FI$, let us consider in greater detail the different detection events possible.
For the system under consideration, there are only two possible cases: we can measure either a coincidence event or an \textit{exclusive} single-photon event i.e., a single click in either one of the two detectors, without a coincidence. 
If we measure a coincidence event, we do not gain any information on which stage generated the pair. 
On the other hand, if we measure an \textit{exclusive} single-photon event, we know that the photon must have been generated in the first stage because internal losses can only affect the pairs generated in the first stage. 
Knowing which stage generated the photon pair breaks the coherence of the system \cite{Luis1996} and thus destroys the interference between the two stages --- often referred to as \textit{which-crystal} interference \cite{Hochrainer2021, Luo2021}.
This is analogous to what happens in Young's double slit experiment, where knowledge of which path the photon has taken erases the interference pattern on the screen.
Therefore, all the events where one knows exactly where the photon pair was generated cannot exhibit any interference pattern, and thus they carry no phase information.
Since this occurs only in the single arms, this leads to the condition $F_{CC}^{\max}\geq F_{A/B}^{\max}$.

The second interesting feature that can be observed in Figure \ref{fig:internal_losses} is that the $FI$ of both singles and coincidences increases proportionally to $g_2^2$ until it reaches a saturation level which depends on the internal transmission of the system and on the number of resources used to probe the sample, $g_1^2$.
Even more interestingly, $\ficcmax$ saturates abruptly to the value of $4g_1^2T_AT_B$ when the coincidences reach maximum visibility for loss-balanced gains, as shown by the round markers in the plot.

To understand the reason behind the observed behaviour of $\ficcmax$, we begin by noting that the interference pattern of the coincidence clicks in Eq.~\eqref{subeq:pcc} arises due to the interference of two probabilities, one with amplitude $g_1\sqrt{T_AT_B}$ and one with amplitude $g_2$. 
These correspond to the probability of detecting a photon pair that was generated in the first or in the second stage, respectively.
Since the amplitude of the first probability is fixed to $g_1\sqrt{T_AT_B}$ three main operating regimes for the interferometer can be identified, depending on the magnitude of $g_2$.

If $g_2<g_1\sqrt{T_AT_B}$, then the probability of measuring a pair generated in the second stage is too small compared to the one of the first stage. 
This results in a poor interference pattern ($V_{CC}<1$), meaning that it is not possible to readout all the information available inside the interferometer.
As the gain of the second stage is increased, the two probabilities start to become comparable in magnitude, i.e., a detection event has similar likelihood of coming from a pair generated in the first or in the second stage.
This is why, in this regime, $\ficcmax$ depends on $g_2^2$: we can vary the amount of information retrieved by changing probability of generating photon pairs in the second stage.

As the interferometer reaches loss-balanced gains ($g_2=g_1\sqrt{T_AT_B}$), the two probabilities have the same amplitude, so perfect visibility $V_{CC}=1$ is reached. 
This means that one can extract all the information available inside the interferometer. 
Thus, in a loss-balanced interferometer, maximum $\ficcmax$ is reached. 
From Eq.~\eqref{subeq:FCCmax}, one can notice that, in this case, $\ficcmax$ is proportional to $g_1^2T_AT_B$.
The reason is that the total number of pairs that carry phase information is limited to $g_1^2T_AT_B$, which is the number of pair that is generated in the first stage and survive the internal losses.

Further increasing the gain $g_2$ does not allow one to retrieve more information since this is limited by the pairs generated in the first stage.
This is the reason why the coincidences saturate at a value that is proportional to $g_1^2T_AT_B$.
The only effect of increasing $g_2$ is to reduce the visibility $V_{CC}$ as the extra events generated in the second stage constitute only a noise floor.

Differently from the coincidences, in Figure \ref{fig:internal_losses} one can see that the $FI$ of the singles saturates later than $\ficcmax$ and with a flatter slope. 
The reason is that, as we have discussed above, the \textit{exclusive} single-photon events carry no information about the internal phase of the interferometer and thus pollute the interference signal.
In fact, since there are no exclusive-single-photon events in the case of perfect internal transmission, the behaviour of the singles is identical to the one of the coincidences, i.e. $\fiamax=\fibmax=\ficcmax$, as can be derived from Eq.~\eqref{eq:fisher_info}.

The analytic description discussed so far is based on the approximation that multiphoton components are not present, such that click detectors are sufficient to accurately reconstruct the state generated inside the interferometer. 
However, it is not yet clear until when this approximation holds true. 
Moreover, it is interesting to see what happens to the $FI$ obtained with binary detectors, when the state generated by the interferometer has non-negligible higher order components. 
For these reasons, we perform numerical simulations using the Python package QuTiP \footnote{The simulation scripts are available upon request.}.
The simulations are performed in the photon-number basis with a Hilbert space size of 10 and consider $g_1 = 0.05$ and $g_2\in [10^{-3}, 1]$. 
For an ideal, lossless system, this corresponds to generating at most a mean photon pair number $\expval{n_{pairs}}\in [0.01, 1.57]$ after the second stage.
Finally, three different sets of transmissions $(T_A,\,T_B)$ have been randomly generated, corresponding to low, moderate and high internal losses. The values of the transmissions are shown in Figure \ref{fig:internal_losses}.

The results of the simulation are shown by the dashed lines in Figure \ref{fig:internal_losses}.
From the figure, one can see that our analytic theory becomes inaccurate when $g_2 \approx 0.2$, regardless of the values of the internal transmissions.
In these conditions, the mean photon pair number generated at the output of the interferometer is mainly determined by the second stage, which is not influenced by the internal losses, and lies in the range $\expval{n_{pairs}}\sim 0.045-0.063$ --- depending on the internal transmissions of the system. 
For such high mean photon pair number, the multiphoton events already constitute $\sim 5\%$ of the total number of single-photon events, and therefore they cannot be neglected, which explains the deviations from the analytic theory.
These results are not specific to the values of gains and transmission chosen for the simulations but stem from the observation that multiphoton components become non-negligible when $\expval{n_{pair}}$ reaches values around 0.05.

\subsection{Imperfect detection}
\label{subsec:imperfect_detection}
So far we have seen that, in the case of perfect detection, it is always better to measure coincidences between the output modes, since their $FI$ is always the highest.  
When detection losses are considered, the situation changes significantly.
In fact, detection losses affects the coincidences much more than the singles, as can be seen from the click probabilities in Eq.~\eqref{eq:click_prob}: the singles depends only on $\eta_{A}$ or $\eta_B$, while the coincidences depend on their product $\eta_A\eta_B$.
This observation implies that, in the presence of external losses, the singles can achieve the absolute maximum $FI$ in an SU(1,1) interferometer, which is limited by the maximum detection efficiency $\eta_{\max} = \max\{\eta_A,\eta_B\}$ and is given by
\begin{equation}
    FI^{\max}_{SU(1,1)} = 4\eta_{\max}g_1^2T_AT_B.
    \label{eq:upper_bound_su11}
\end{equation}

\begin{figure}[tbp]
    \centering
    \includegraphics[width =\columnwidth]{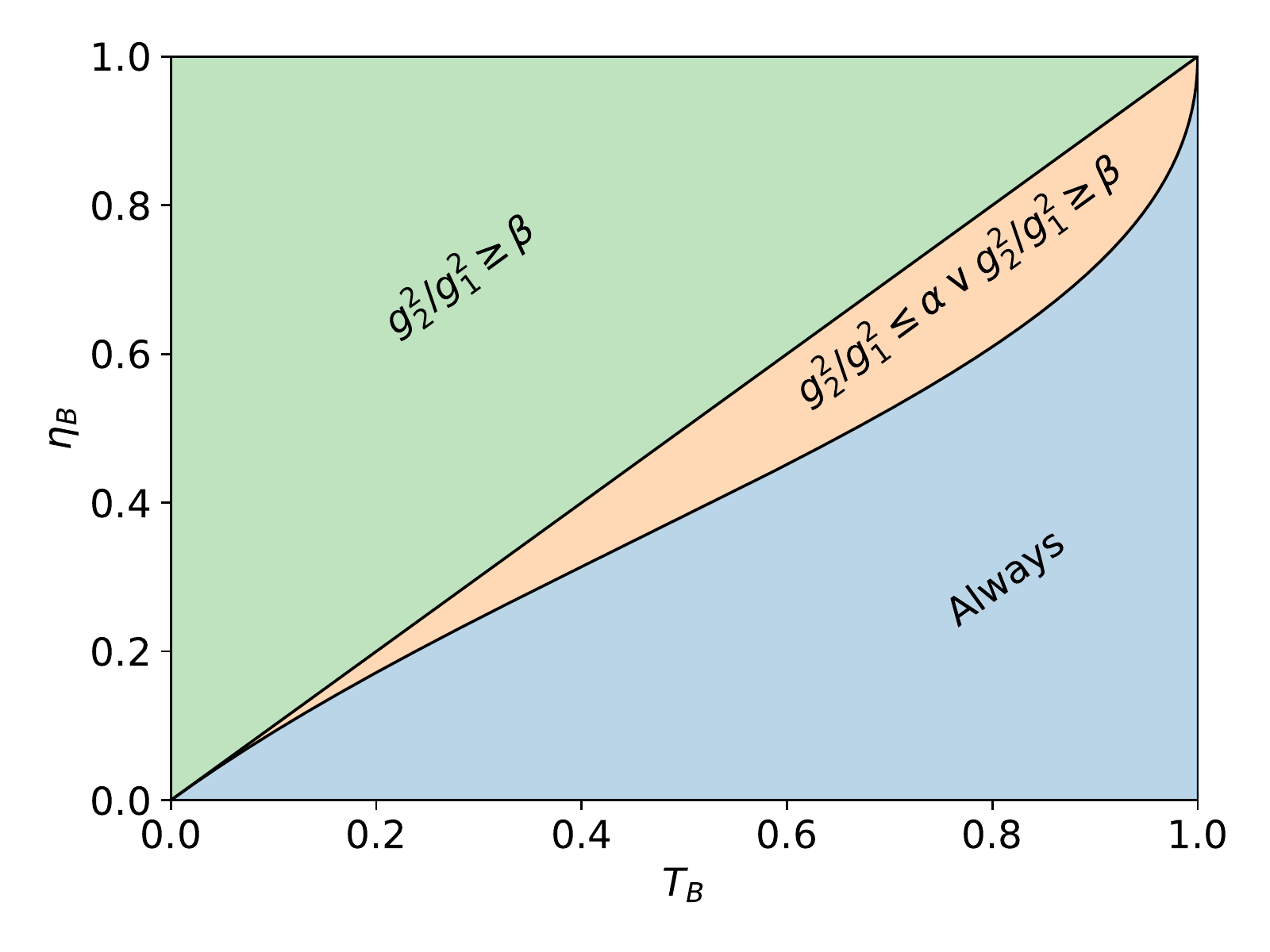}
    \caption{Visualization of the conditions in Eq.~\eqref{eq:conditions_imperfect_detection}, representing the experimental conditions where is advantageous to measure the singles in mode A instead of coincidences ($FI^{\max}_{CC}\geq FI^{\max}_{A/B}$), depending on the internal transmission $T_B$ and the detection efficiency $\eta_B$ of the other mode. 
    The boxes report the conditions on the gains $g_1$ and $g_2$ where the given inequality holds true.
    The values of $\alpha$ and $\beta$ are reported in Eq.~\eqref{eq:bounds}.
    Due to symmetry arguments, the same results hold true if we consider $\fibmax$.}
    \label{fig:boundaries_det_eff}
\end{figure}

For a better understanding of the relation between the $FI$ of the singles and the one of the coincidences in the case of imperfect detection, we compare in the following $\ficcmax$ and $\fiamax$ --- note that the choice of mode A is arbitrary as mode A and mode B are interchangeable due to the vacuum seeding in both modes.
With straightforward but rather lengthy calculations, it can be shown that the $\fiamax$ is greater than $\ficcmax$ in the following scenarios:

\begin{equation}
    \fiamax\geq \ficcmax \iff 
    \begin{cases}
    g_2^2/g_1^2 \geq \beta & \text{if } T_B<\eta_B,\\[2ex]
    g_2^2/g_1^2\leq \alpha \text{ or } g_2^2/g_1^2\geq \beta & \text{if } \eta_B\leq T_B \leq \frac{\eta_B}{1-\eta_B+\eta_B^2},\\[2ex]
    \text{Always} &\text{if } T_B > \frac{\eta_B}{1-\eta_B+\eta_B^2},
    \end{cases}
\label{eq:conditions_imperfect_detection}
\end{equation}

with
\begin{equation}
    \alpha =\frac{T_A\pqty{T_B-\eta_B}}{\eta_B\pqty{1-\eta_B}}\quad \text{and}\quad    \beta = \eta_B T_A \frac{1-\eta_B T_B}{1-\eta_B}.
    \label{eq:bounds}
\end{equation}
An intuitive representation of the conditions in Eq.~\eqref{eq:conditions_imperfect_detection} is represented in Figure \ref{fig:boundaries_det_eff}. 
Note that, from symmetry arguments, analogous conditions can be found for $\ficcmax\geq \fibmax$.

The main conclusion of Eq. \eqref{eq:conditions_imperfect_detection} is that, in the presence of detection losses, the $FI$ of the singles eventually becomes greater than $\ficcmax$ if the gain $g_2$ is sufficiently high. 
Moreover, in the presence of low internal losses and relatively high detection losses, measurement of the singles always outperforms the coincidence measurements.
This result is quite interesting, given that we have seen that the phase information is encoded in the photon pairs and not in the singles.
What happens is that, thanks to the photon-number correlations of states generated in the nonlinear stages, the singles at the output of the interferometer still retain the information carried by the interference of the photon pairs, which can be extracted even in the case of low detection efficiencies.

Similar to the case of ideal detection, we have simulated the effect of imperfect detection for a number of systems to investigate when our analytic theory starts to become invalid. 
For simplicity, the simulations considered an interferometer with $T_A = 20\%$ and $T_B = 22\%$, corresponding to case (3) in Figure \ref{fig:internal_losses}, and with varying but equal detection efficiencies in both arms, i.e. $\eta_A=\eta_B = \eta_{detection}$. 

\begin{figure}[tbp]
    \centering
    \includegraphics[width =\columnwidth]{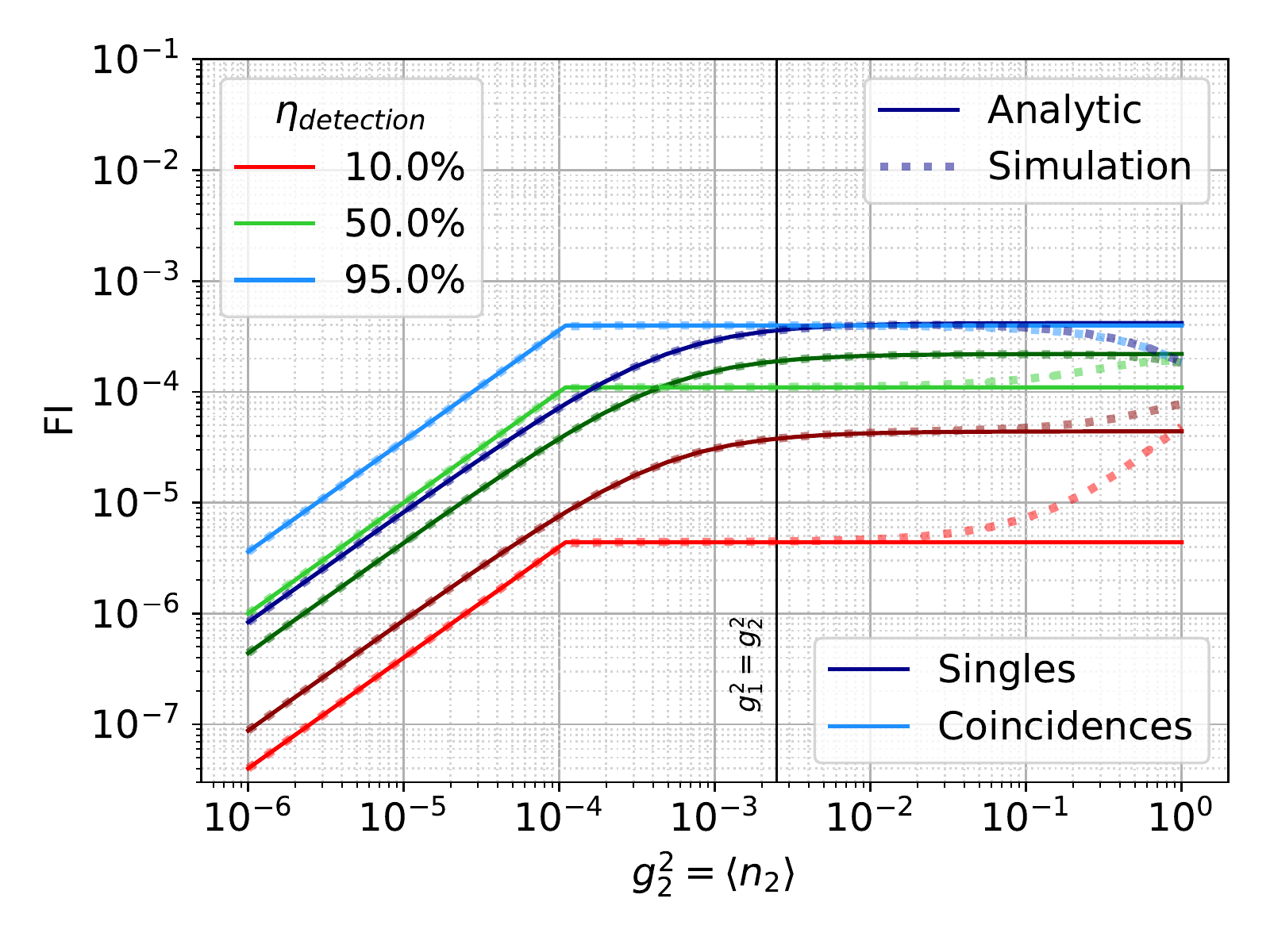}
    \caption{$FI$ in a lossy SU(1,1) interferometer with imperfect detection. The gain of the first stage is set to $g_1 = 0.05$, and the internal transmissions are $T_1 = 20\%$ and $T_2 = 22\%$. 
    The solid colour lines correspond to the analytic theory while the dashed colour lines correspond to the simulations (more details in the text). 
    The darker lines correspond to the $FI$ of the singles while the lighter lines correspond to the $FI$ of the coincidences. 
    For $g_2^2>0.01$, multiphoton components cause a deviation from the constant $FI$ predicted by the analytic theory, especially for low detection efficiency.}
    \label{fig:det_efficiency}
\end{figure}

The results of the simulations are displayed in Figure \ref{fig:det_efficiency}.
In agreement with our analytic theory, one can see that, when the detection efficiency is lower than the internal transmission of the system ($\eta = 10\%$), the $FI$ of the singles is always above the $FI$ of the coincidence.
As the detection efficiency increases, the $FI$ of the coincidences becomes higher than the one of the single, for $g_2/g_1\leq \beta$.

Finally, the figure shows that the deviation between the analytic theory and the numerical simulations now depends on the detection efficiency $\eta$.
This is somewhat expected since now the photon statistics of the state that exits the interferometer can be affected by the presence of detection losses.
Interestingly, for low detection efficiencies, we observe higher $FI$ for both singles and coincidences than predicted by the analytic theory, in contrast with what observed in the cases of high detection efficiency.
We attribute this to the fact that having multiphoton components now helps overcoming the high detection losses.

\subsection{Optimization of a low-gain, lossy SU(1,1) interferometer}
The results presented so far provide different strategies for maximizing the $FI$ measured using a low-gain, lossy SU(1,1) interferometer, assuming the gain $g_1$ is constrained. 
We can consider two main scenarios: in the first one, we are allowed to freely choose the gain $g_2$ of the second stage; in the second one, we are restricted to a maximum level $g_2$, e.g., due to laser power limitations.

In the first case, the results of the last section show that one can simply set the interferometer such that $g_2\gg g_1$, as we have demonstrated that, for this condition, the $FI$ of the singles is as good as the $FI$ of the coincidences --- or even better, in the presence of detection losses.

In the second case, a more refined analysis of the interferometer is necessary. 
The first step is to characterise the detection losses of the setup. 
This can be done simply by pumping only the second stage of the interferometer and measuring the Klyshko efficiencies \cite{Klyshko_1980}, which correspond to the detection losses.
Then, the first stage is pumped with gain $g_1$ while the gain $g_2$ is scanned and the visibilities $V_A$, $V_B$ and $V_{CC}$ are monitored.
In this way, it is possible to use Eq.~\eqref{eq:visibilities} to fit the values of the internal transmissions $T_A$ and $T_B$. 
Finally, now that both the internal and the detection losses are known, one can use Eq. \eqref{eq:conditions_imperfect_detection} to determine the optimal $g_2$. 
As general a rule of the thumb, if the internal transmission is higher than the detection efficiency, it is advised to measure the singles; if, on the other hand, the internal transmission is lower than the detection efficiency, then it is better to measure the coincidences when $g_2^2\leq g_1^2$, and the singles when $g_2^2\gg g_1^2$.

Interestingly, there is a second way to determine the internal transmissions of the system. 
At loss-balanced gains ($g_2^2 = g_1^2T_AT_B$), one can use Eqs.~\eqref{subeq:visa} and \eqref{subeq:visb} to easily calculate the transmissions of modes A and B inside the interferometer as
\begin{subequations}
\begin{align}
    T_A &= \frac{V_B}{2-V_B},\\
    T_B &= \frac{V_A}{2-V_A}.
\end{align}
\end{subequations}
This result is quite striking: it is possible to use a loss-balanced SU(1,1) interferometer to estimate the losses inside the interferometer by simply measuring the visibility of the singles.

\section{Comparison SU(2) and SU(1,1) interferometers}
\label{sec:comparison}
In the previous section, we have discussed the properties of lossy, low-gain SU(1,1) interferometers and how internal and external losses impact on the $FI$ of the measured singles and coincidences.
However, it is not obvious yet when is advantageous to implement these interferometers instead of classical MZIs.
In other words, we have not discussed in which experimental conditions SU(1,1) interferometers can beat the SQL.
To answer this question, in this last part of our work, we investigate in which context it is beneficial to set up an SU(1,1) interferometer, as opposed to a classical MZI --- also known as SU(2) interferometer.

A fair comparison between the two interferometers can be performed by analysing their $FIs$ under the assumption that they are using the same amount of resources.
This is easier said than done since the choice of which are the resources to be counted is not unique.
Possible choices in the SU(1,1) interferometer are the total number of photons generated at the output of the interferometer, the number of photons generated in the first stage, the number of photon pairs generated in the first stage, or the number of photons in the pump field that is used to generate the photon pairs.

In the following, we decide to consider as resources the numbers of photons passing through the sample. 
This choice has both a practical and a theoretical motivation.
From an experimental perspective, we are quite often interested in keeping track of how many photons interact with the sample under test, in particular when it can be damaged.
From a theoretical perspective, this choice is motivated by the fact that, in an SU(1,1) interferometer, the $FI$ is ultimately limited by the number of photon pairs generated by the first stage (see Eq.~\eqref{eq:upper_bound_su11}), which also corresponds to the number of photons that pass through the sample.
For this reason, we compare the $FI$ of classical SU(2) and SU(1,1) interferometers assuming that the same number $\expval{n}=g_1^2\ll 1$ of photons is used to probe the phase element. 
In the case of an SU(1,1) interferometer, this corresponds to setting the gain of the first stage to $g_1^2 = \expval{n}$.
In the case of a classical SU(2) interferometer, this corresponds to injecting the MZI with a coherent state $\ket{\alpha}$ having $\abs{\alpha}^2 = 2\expval{n}$.
Note that here we consider $g_1$ to be fixed, since we are not interested here in the scaling properties of the two systems as a function of the mean photon number $\expval{n}$ inside the interferometer as both systems scale with $\sqrt{\expval{n}}$ for $\expval{n}\ll 1$ \cite{Sparaciari2016, Florez2018}.

\begin{figure}[tbp]
    \centering
    \includegraphics[width=0.8\columnwidth]{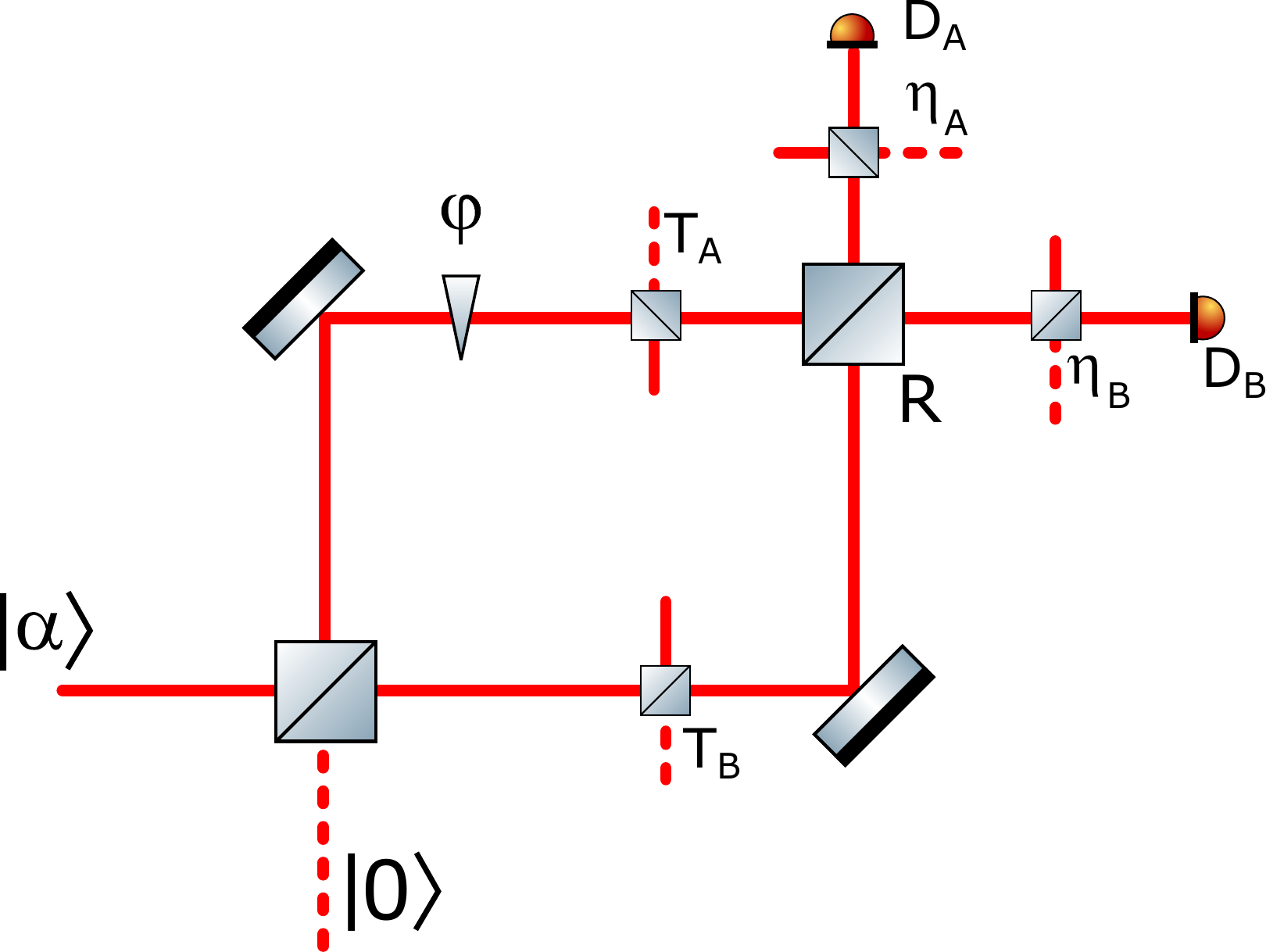}
    \caption{Sketch of the lossy SU(2) interferometer used in the comparison with the SU(1,1) interferometer of Figure \ref{fig:su11_sketch}.}
    \label{fig:MZI}
\end{figure}

To compare SU(1,1) and SU(2) interferometers, we need to describe the SU(2) interferometer in the same framework that was used in the previous sections.
The system under consideration is the MZI sketched in Figure \ref{fig:MZI}.
It consists of a 50:50 input beam splitter, an output beam splitter with variable reflectivity $R$. 
The internal transmissions and detection efficiencies are modelled similarly to the SU(1,1) interferometers, with beam splitters with transmissions $T_{A/B}$ and $\eta_{A/B}$, respectively.
We consider the state $\ket{\alpha, 0}$ as input, where a coherent state with amplitude $\abs{\alpha}^2 = 2\expval{n} = 2g_1^2\ll 1$ is injected in one input arm and vacuum is present at the other.
For the system described, the probability of measuring singles in mode $A/B$ is given by
\begin{align}
    p_{A/B} = & \eta_{A/B}g_1^2\left[\pqty{1-R}T_{A/B} + RT_{B/A}\right.\notag\\
    & \left.-2\sqrt{\pqty{1-R}RT_AT_B}\cos\pqty{\phi}\right].
\end{align}
We do not consider here the probability of coincidence events between the output ports of the MZI because it tends to zero, since we are injecting a state with $\expval{n}\ll1$.

Following the same strategy as in Section \ref{sec:analy_theory}, it is possible to calculate the maximum $FI$ for the SU(2) interferometer. 
If $\eta_{\max}=\max\qty{\eta_A, \eta_B}$ is the highest detection efficiency of the setup, the maximum $FI$ of the system under study is given by
\begin{subequations}
\begin{align}
    FI^{\max}_{SU(2)} &= 2\eta_{\max}\abs{\alpha}^2 \frac{T_AT_B}{T_A+T_B}\label{subeq:fi_su2_coh}\\
    &= 4\eta_{\max}g_1^2 \frac{T_AT_B}{T_A+T_B}.
    \label{subeq:fi_su2}
\end{align}
\label{eq:fi_su2}
\end{subequations}

One can use the upper bounds of Eqs.~\eqref{eq:upper_bound_su11} and \eqref{eq:fi_su2} to quickly compare different SU(1,1) and SU(2) systems.
In fact, these equations can be used to answer two interesting questions: for the classical system, how many more photons need to interact with the sample, to achieve the same $FI$ of an identical SU(1,1) interferometer? 
And, for the same number of resources, how high should the detection efficiency of a classical system be, in order to achieve the same performance of an identical SU(1,1) interferometer?
To answer the first question, we consider  Eqs.~\eqref{eq:upper_bound_su11} and \eqref{subeq:fi_su2_coh} to be equal and assume that internal transmissions and detection efficiencies are the same. 
In this way, one derives the mean photon number impinging on the phase element in an SU(2) interferometer must be $T_A+T_B$ times higher than the mean photon pairs generated by the first stage in an SU(1,1) interferometer. 
Similarly, to answer the second question, we consider Eqs.~\eqref{eq:upper_bound_su11} and \eqref{subeq:fi_su2} to be equal and assume that the internal transmissions of the two systems are the same.
Solving for the detection efficiency, one can derive that an SU(2) interferometer needs a detection efficiency that is $T_A+T_B$ times higher than the one of an SU(1,1) interferometer in order to achieve the same $FI$.
As an example, considering an SU(1,1) and an SU(2) interferometer with realistic internal transmissions $T_A=T_B=60\%$, the sample inside the SU(2) interferometer needs to interact with 20\% more photons, on average, to achieve the same $FI$.
For the same interferometers, if the mean photon number sampling the phase element is the same, the SU(2) interferometer needs a detection efficiency 20\% higher than an SU(1,1) interferometer.

The result of this analysis shows that the sum of the internal transmission of an SU(1,1) interferometer is the figure of merit to compare it to a classical SU(2) interferometer.
This means that SU(1,1) interferometers characterized by low losses and high-quality detectors can gain up to a factor of 2 over the SQL.

As we have seen from the examples above, it is appears that one of the main metrics of comparison between SU(1,1) and SU(2) interferometers is the sum of the internal transmissions $T_A+T_B$. 
Indeed, using Eqs.~\eqref{eq:upper_bound_su11} and \eqref{eq:fi_su2} to compare the $FI$ performance of SU(1,1) and the SU(2) interferometers with the same internal transmissions $T_{A/B}$ and the same maximum detection efficiency $\eta_{\max}$ in the asymptotic limit $g_2^2\gg g_1^2$, we discover that the SU(1,1) interferometer can outperform the SU(2) systems when 
\begin{equation}
    FI^{\max}_{SU(1,1)}>FI^{\max}_{SU(2)} \iff T_A+T_B>1. \label{eq:conditional_advantage}
\end{equation}
We call this condition \textit{asymptotic conditional advantage} because it was derived in the asymptotic limit and because it is conditioned to both system having the same internal losses.

One can also perform a stricter comparison and consider a lossy SU(1,1) and a lossless SU(2) interferometer.
In this case, the SU(1,1) can still outperform the classical system when
when
\begin{equation}
    FI^{\max}_{SU(1,1)}>FI^{\max}_{\text{ideal } SU(2)} \iff 2\eta_{\max}T_A T_B >1.
    \label{eq:unconditional_advantage}
\end{equation}
We refer to this condition as \textit{asymptotic unconditional advantage} because we have posed no constraints on the classical system.

The conditions \eqref{eq:conditional_advantage} and \eqref{eq:unconditional_advantage} are very handy rules of thumb to assess whether the SU(1,1) interferometer can, in general, beat classical systems. 
For example, assuming equal transmissions $T_A=T_B=T$ and perfect detection, one can see that asymptotic conditional advantage can be achieved with internal transmission $T>50\%$ while asymptotic unconditional advantage can be achieved with $T>71\%$.

The asymptotic conditions are useful to determine whether the SU(1,1) interferometer can, in principle, beat the SQL.
However, they do not provide direct information of what are the experimental conditions that allow one to actually beat the SQL. 
The answer to this question can be obtained with a direct comparison of Eqs.~\eqref{eq:fisher_info} and Eq.~\eqref{eq:fi_su2}.
In this way, we calculate the minimum gain ratio $g_2^2/g_1^2$ necessary to achieve the conditional/unconditional advantage, when considering both the internal and the external transmissions of the interferometers. 
The results of this comparison are quite involved and are reported in Table \ref{tab:gains}.

\begin{table*}[tb]
\caption{\label{tab:gains}
Table containing the minimum gain ratio $g_2^2/g_1^2$ that allows conditional/unconditional advantage of an SU(1,1) interferometer with respect to a classical SU(2) interferometer. $T_{A/B}$ are the internal transmissions for mode $A/B$, $\eta_{A/B}$ are the detection efficiencies for mode $A/B$, and $\eta_{\max}$ is the highest detection efficiency of the classical SU(2) interferometer.
For simplicity, only the condition for the singles in mode A are reported. From these, the results for the singles in mode B can be obtained simply by swapping the labels A and B.}

\scriptsize
    \centering
    \begin{tabular}{lll}
    Advantage & Singles (Mode A) & Coincidences \\
    \hline\\
    Conditional &  
    $g_2^2/g_1^2\geq\frac{T_A\eta_{\max}\bqty{\eta_{\max} T_B-\eta_A\pqty{T_A+T_B}}  }{\eta_A\pqty{T_A+T_B}\bqty{\eta_{\max}-\eta_A\pqty{T_A+T_B}}}$, if $T_A+T_B>\frac{\eta_{\max}}{\eta_A} $ & 
    $g_2^2/g_1^2\geq\frac{\eta_{\max}T_A T_B}{\eta_A\eta_B\pqty{T_A+T_B}}$, if $T_A+T_B>\frac{\eta_{\max}}{\eta_A\eta_B}$\\
    \vspace{0.2cm}\\
    Unconditional & 
    $g_2^2/g_1^2\geq\frac{\eta_{\max}\pqty{\eta_{\max}-2T_A\eta_A } }{2\eta_A \pqty{\eta_{\max} - 2 T_A T_B \eta_A } }$, if $T_AT_B > \frac{\eta_{\max}}{2\eta_A}$ &
    $g_2^2/g_1^2\geq\frac{\eta_{\max} }{2\eta_A\eta_B}$, if $T_AT_B > \frac{\eta_{\max}}{2\eta_A\eta_B}$
    \end{tabular}
\end{table*}

As a final comment, it is clear that also other systems can beat the SQL in the regime of $\expval{n}<1$ considered here, e.g., NOON state interferometry \cite{Slussarenko2017}.
However, the analysis presented here highlights that SU(1,1) interferometer can perform well even in the presence of relatively high internal losses, which typically constitutes a problem for other quantum-enhanced systems.

\section{Conclusions}
In this paper, we derived and discussed the classical $FI$ of lossy SU(1,1) interferometers when states with $\expval{n}<1$ are generated inside the setup and when utilizing click detectors. 
We have shown that, in the case of lossless detection, coincidence measurement always performs better than measuring a single arm and that the classical $FI$ of the coincidence measurement saturates to the highest possible value when the interferometer is loss-balanced. 
Interestingly, when detection losses are considered, we found that measuring the single events can yield higher $FI$ than coincidences.

Finally, we discussed the conditions where lossy SU(1,1) interferometers outperform their classical SU(2) counterparts, in the limits of low photon numbers inside the setup.
We discovered that the main figure of merit of an SU(1,1) interferometer is the sum of the internal transmissions $T_A+T_B$. 
When this parameter is above 1, the SU(1,1) interferometer can be more efficient than a similar SU(2) interferometer.   

The results presented in this paper are of fundamental importance for the understanding of SU(1,1) interferometers, in particular in the regime of low mean photon numbers. 
Our findings show the power of photon-number correlations and how they can help overcome the presence of internal losses inside the interferometer. 

These findings are of further significance for the implementation of SU(1,1) interferometers for the investigation of lossy and/or photosensitive samples and thus will likely impact the future development of these interferometers for practical applications.

\section{Acknowledgments}
M. Santandrea thanks L.~Serino and P.~Folge for support during the development of the analytic models in the paper.
M. Santandrea acknowledges financial support from the Federal Ministry of Education and Research (BMBF) via the grant agreement no. 13N15065 (MiLiQuant).
K. H. L and H. H. acknowledge financial support through the Deutsche Forschungsgemeinschaft (DFG -- German Research Foundation) (NO. 231447078 -- TRR 142 C02 and A09).
J. S. acknowledges financial support from the Deutsche Forschungsgemeinschaft (DFG, German Research Foundation) through the Collaborative Research Center TRR 142 (Project No. 231447078, project C10).

%%%%%%%%%%%%%%%%%%%%%%% References %%%%%%%%%%%%%%%%%%%%%%%%%

\section*{References}

\providecommand{\newblock}{}

\newpage
\appendix
\section{Calculation of the singles and coincidence probabilities}
\label{supp:sec:su11calc}

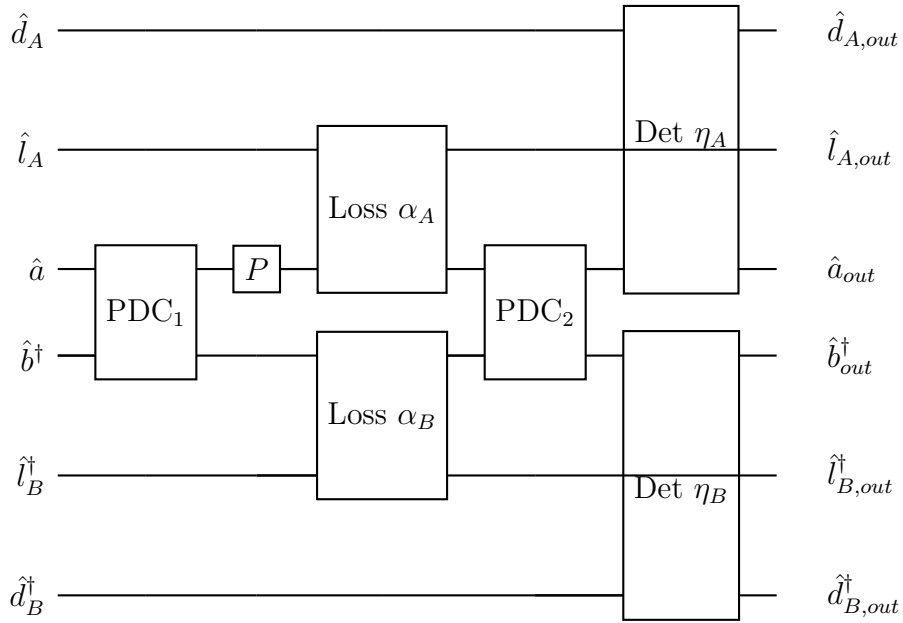
\begin{figure}[bp]
    \centering
    \begin{quantikz}[transparent]
        \lstick{$\hat{d}_{A}$}  & \qw & \qw & \qw & \qw & \gate[3,label style={yshift=0.2cm}]{\text{Det }\eta_A} & \qw& \rstick{$\hat{d}_{A, out}$}\\
        \lstick{$\hat{l}_{A}$} & \qw & \qw & \gate[wires=2]{\text{Loss }\alpha_A} & \qw & \linethrough & \qw& \rstick{$\hat{l}_{A, out}$}\\
        \lstick{$\hat{a}$} & \gate[wires=2]{\text{PDC}_1} & \gate{P} & \qw & \gate[wires=2]{\text{PDC}_2} & \qw &  \qw&\rstick{$\hat{a}_{out}$}\\
        \lstick{$\hat{b}^\dag$} & \qw & \qw & \gate[wires=2]{\text{Loss }\alpha_B} & \qw & \gate[3,label style={yshift=-0.2cm}]{\text{Det }\eta_B} & \qw& \rstick{$\hat{b}_{out}^\dag$}\\
        \lstick{$\hat{l}_B^\dag$} & \qw & \qw & \qw & \qw & \linethrough & \qw & \rstick{$\hat{l}^\dag_{B, out}$}\\
        \lstick{$\hat{d}_B^\dag$}  & \qw & \qw & \qw & \qw & \qw & \qw & \rstick{$\hat{d}^\dag_{B, out}$}\\
    \end{quantikz}
    \caption{Schematics for the SU(1,1) interferometer with losses and non-ideal detection efficiency.}
    \label{fig:scheme}
\end{figure}

The SU(1,1) interferometer shown in Figure 1 of the main paper can be described in a quantum-circuit model, as shown in Figure \ref{fig:scheme} for bosonic annihilation and creation operators. 
In the following, we indicate with  $\hat a$ and $\hat b$ the input modes of the nonlinear interferometer, and with $\hat a_{out}$ and $\hat b_{out}$ the respective output modes. 
Internal losses are modelled using independent beam splitters with (power) reflectivity of $\alpha_{A,B}= 1-T_{A,B}$. The input (output) vacuum modes of these beam splitters are labelled $\hat l_{A}$ and $\hat l_B$ ($\hat l_{A, out}$ and $\hat l_{B, out}$).
Detection losses are modelled in a similar way, with two independent beam splitters having (power) reflectivity of $1-\eta_{A, B}$. The input (output) vacuum modes of these beam splitters are labelled $\hat d_{A}$ and $\hat d_B$ ($\hat d_{A, out}$ and $\hat d_{B, out}$).
The matrices describing these processes are given in Eqs. \eqref{eq:PDC_1}--\eqref{eq:detectors}, using the shorthand notation $C_{j} = \cosh{g_{j}}$ and $S_{j} = \sinh{g_{j}}$, where $j\in\{1,2\}$.

The involved processes are given in terms of the following matrices:
\begin{align}
\text{PDC}_1 &=     \pqty{
    \begin{array}{cccccc}
         1 & 0 & 0 & 0 & 0 & 0 \\
         0 & 1 & 0 & 0 & 0 & 0 \\
         0 & 0 & C_1 & S_1 & 0 & 0 \\
         0 & 0 & S_1 & C_1 & 0 & 0 \\
         0 & 0 & 0 & 0 & 1 & 0 \\
         0 & 0 & 0 & 0 & 0 & 1
    \end{array}\label{eq:PDC_1}
    },\\
\text{PDC}_2 &=     \pqty{
    \begin{array}{cccccc}
         1 & 0 & 0 & 0 & 0 & 0 \\
         0 & 1 & 0 & 0 & 0 & 0 \\
         0 & 0 & C_2 & e^{-i \theta } S_2 & 0 & 0 \\
         0 & 0 & e^{i \theta } S_2 & C_2 & 0 & 0 \\
         0 & 0 & 0 & 0 & 1 & 0 \\
         0 & 0 & 0 & 0 & 0 & 1
    \end{array}
    },\\
\text{P} &=     \pqty{
    \begin{array}{cccccc}
         1 & 0 & 0 & 0 & 0 & 0 \\
         0 & 1 & 0 & 0 & 0 & 0 \\
         0 & 0 & e^{i\phi} & 0 & 0 & 0 \\
         0 & 0 & 0 & 1 & 0 & 0 \\
         0 & 0 & 0 & 0 & 1 & 0 \\
         0 & 0 & 0 & 0 & 0 & 1
    \end{array}
    },\\
\text{Loss} &=     \pqty{
   \begin{array}{cccccc}
     1 & 0 & 0 & 0 & 0 & 0 \\
     0 & \sqrt{1-\alpha_A} & \sqrt{\alpha_A} & 0 & 0 & 0 \\
     0 & -\sqrt{\alpha_A} & \sqrt{1-\alpha_A} & 0 & 0 & 0 \\
     0 & 0 & 0 & \sqrt{1-\alpha_B} & -\sqrt{\alpha_B} & 0 \\
     0 & 0 & 0 & \sqrt{\alpha_B} & \sqrt{1-\alpha_B} & 0 \\     
     0 & 0 & 0 & 0 & 0 & 1
    \end{array}
    },\\
    \text{Det} &=     \pqty{
    \begin{array}{cccccc}
         \sqrt{\eta_A} & 0 & \sqrt{1-\eta_A} & 0 & 0 & 0 \\
         0 & 1 & 0 & 0 & 0 & 0 \\
         -\sqrt{1-\eta_A} & 0 & \sqrt{\eta_A} & 0 & 0 & 0 \\
         0 & 0 & 0 & \sqrt{\eta_B} & 0 & -\sqrt{1-\eta_B} \\
         0 & 0 & 0 & 0 & 1 & 0 \\
         0 & 0 & 0 & \sqrt{1-\eta_B} & 0 & \sqrt{\eta_B}
    \end{array}
    }\label{eq:detectors},
\end{align}
whose rows and columns are sorted as $\hat d_A$, $\hat l_A$, $\hat a$, $\hat b^\dag$, $\hat l_B^\dag$, and $\hat d_B^\dag$, according to Figure \ref{fig:scheme}.
Please note that, for the sake of convenience, the first and last three entries respectively act on annihilation and creation operators.
Furthermore, $\phi$ and $\theta$ denote phases, and $\alpha_j$ and $\eta_j$ for mode $j\in\{A,B\}$ define propagation losses and detection efficiencies, respectively.
The composed transfer matrix between the input and output modes is thus given by 
\begin{equation*}
\text{T} = \text{Det}\cdot\text{PDC}_2\cdot \text{Loss}\cdot \text{P}\cdot\text{PDC}_1,    
\end{equation*}
which yields
\begin{equation}
    \text{T} =\pqty{
    \begin{smallmatrix}
     \sqrt{\eta_A} & -C_2\sqrt{\alpha_A} \sqrt{1-\eta_A} & \mathcal{A} \sqrt{1-\eta_A}
       & \mathcal{B} \sqrt{1-\eta_A} & -e^{-i\theta}S_2\sqrt{\alpha_B}
       \sqrt{1-\eta_A} & 0 \\
     0 & \sqrt{1-\alpha_A} & C_1 e^{i \phi } \sqrt{\alpha_A} & e^{i \phi }
       S_1 \sqrt{\alpha_A} & 0 & 0 \\
     -\sqrt{1-\eta_A} & -C_2\sqrt{\alpha_A} \sqrt{\eta_A} & \mathcal{A} \sqrt{\eta_A}
       & \mathcal{B} \sqrt{\eta_A} & -e^{-i\theta}S_2\sqrt{\alpha_B}
       \sqrt{\eta_A} & 0 \\
     0 & -e^{i \theta} S_2 \sqrt{\alpha_A}\sqrt{\eta_B} & \mathcal{C}
       \sqrt{\eta_B} & \mathcal{D} \sqrt{\eta_B} & -C_2\sqrt{\alpha_B}
       \sqrt{\eta_B} & -\sqrt{1-\eta_B} \\
     0 & 0 & S_1 \sqrt{\alpha_B} & C_1 \sqrt{\alpha_B} &
       \sqrt{1-\alpha_B} & 0 \\
     0 & -e^{i \theta} S_2 \sqrt{\alpha_A} \sqrt{1-\eta_B} & \mathcal{C}
       \sqrt{1-\eta_B} & \mathcal{D} \sqrt{1-\eta_B} & -C_2\sqrt{\alpha_B} \sqrt{1-\eta_B} & \sqrt{\eta_B}
    \end{smallmatrix}
    },
    \label{eq:transfermatrix}
\end{equation}
where%\unsure{The factor $\mathcal{B}$ is different between me and Kai, the position of the sqrt is different.}
%JS check:
\begin{align*}
    \mathcal{A} &= e^{i\phi}\bqty{\sqrt{1-\alpha_A} C_1 C_2 +\sqrt{1-\alpha_B} e^{-i (\theta+\phi)} S_1 S_2},\\
   \mathcal{B} &= e^{i\phi}\bqty{\sqrt{1-\alpha_A}  S_1 C_2 +\sqrt{1-\alpha_B} C_1  S_2 e^{-i (\theta+\phi) }},\\
   \mathcal{C} &= \sqrt{1-\alpha_A} C_1 S_2 e^{i\theta +i \phi } + \sqrt{1-\alpha_B}  S_1C_2,\\
   \mathcal{D} &= \sqrt{1-\alpha_A} S_1 S_2 e^{i
   \theta +i \phi } + \sqrt{1-\alpha_B} C_1 C_2.
\end{align*}
Therefore, the relevant output modes for $\hat{a}$ and $\hat{b}^\dag$ are given by
\begin{subequations}
\begin{align}
    \hat{a}_\text{out} &= -\sqrt{1-\eta_A} \hat{d}_A -C_2\sqrt{1-T_A} \sqrt{\eta_A} \hat{l}_A + \mathcal{A} \sqrt{\eta_A} \hat{a}+ \mathcal{B} \sqrt{\eta_A} \hat{b}^\dag- e^{-i\theta}S_2\sqrt{1-T_B} \sqrt{\eta_A} \hat{l}_B^\dag
    \nonumber\\
    &= D_A \hat{d}_A + L_A \hat{l}_A + A \hat{a}+ B^* \hat{b}^\dag+ L_B^* \hat{l}_B^\dag,\\
    \hat{b}_\text{out}^\dag &=  -e^{i \theta} S_2 \sqrt{1-T_A}\sqrt{\eta_B} \hat{l}_A + \mathcal{C} \sqrt{\eta_B}\hat{a} + \mathcal{D} \sqrt{\eta_B}\hat{b}^\dag - C_2\sqrt{\eta_B} \sqrt{1-T_B}\hat{l}_B^\dag - \sqrt{1-\eta_B}\hat{d}_B^\dag
    \nonumber\\
    &= \tilde{L}_A \hat{l}_A + \tilde{A}\hat{a} + \tilde{B}^*\hat{b}^\dag + \tilde{L}_B^*\hat{l}_B^\dag + D_B^*\hat{d}_B^\dag,
\end{align}
\end{subequations}
where we replaced $\alpha_j = 1-T_j$ ($j\in\{A,B\}$) and indicated the coefficients in \eqref{eq:transfermatrix} with $D_{A/B}$, $L_{A/B}$, $A$ and $B$ for ease of writing.
% and 
% \begin{align*}
%     \mathcal{A} &= e^{i\phi}\bqty{\sqrt{T_A} C_1 C_2 +\sqrt{T_B} e^{-i (\theta+\phi)} S_1 S_2}\\
%   \mathcal{B} &= e^{i\phi}\bqty{\sqrt{T_A}  S_1 C_2 +\sqrt{T_B} C_1  S_2 e^{-i (\theta+\phi) }}\\
%   \mathcal{C} &= \sqrt{T_A} C_1 S_2 e^{i\theta +i \phi } + \sqrt{T_B}  S_1C_2\\
%   \mathcal{D} &= \sqrt{T_A} S_1 S_2 e^{i
%   \theta +i \phi } + \sqrt{T_B} C_1 C_2 
% \end{align*}

Applying the definition of photon-number operators and using multimode vacuum states as inputs, $|\mathrm{vac}\rangle$, the mean photon number for the singles in mode A and B are given by
\begin{align}
\begin{aligned}
    \expval{n_a} =& \eta_A\left[S_1^2C_2^2 T_A+C_1^2S_2^2T_B+S_2^2(1-T_B)\right.\\ &\left.+2S_1S_2C_1C_2\sqrt{T_AT_B}\cos\pqty{\theta+\phi}\right],\\
    \expval{n_b} =& \eta_B\left[S_1^2C_2^2T_B+C_1^2S_2^2T_A+S_2^2\pqty{1-T_A}\right.\\ &\left.+2S_1S_2C_1C_2\sqrt{T_AT_B}\cos\pqty{\theta+\phi}\right].
\end{aligned}
\end{align}
And the coincidences read
\begin{align}
\begin{aligned}
    \expval{n_an_b} =& \expval{\hat{a}_{out}^\dag \hat{a}_{out}\hat{b}_{out}^\dag \hat{b}_{out}}=\\
    =&\bra{\mathrm{vac}}
    \bqty{B\hat{b} +L_B\hat{l}_b} 
    \bqty{D_A \hat{d}_A + L_A \hat{l}_A + A \hat{a}+ B^* \hat{b}^\dag+ L_B^* \hat{l}_B^\dag}\\
    &\times\bqty{\tilde{L}_A \hat{l}_A + \tilde{A}\hat{a} + \tilde{B}^*\hat{b}^\dag + \tilde{L}_B^*\hat{l}_B^\dag + D_B\hat{d}_B^\dag}
    \bqty{\tilde{L}_A^* \hat{l}_A^\dag + \tilde{A}^*\hat{a}^\dag}\ket{\mathrm{vac}}\\
    =&\expval{n_a}\expval{n_b}+B\tilde{B}^*L_A\tilde{L_A}^*+A\tilde{A}^*B\tilde{B}^*+L_A\tilde{L_A}^*L_B\tilde{L_B}^*+A\tilde{A}^*L_B\tilde{L_B}^*\\
    =&\expval{n_a}\expval{n_b}+\pqty{L_A\tilde{L_A}^*+A\tilde{A}^*}\pqty{B\tilde{B}^*+L_B\tilde{L_B}^*}.
\end{aligned}
\end{align}
Considering the low-gain regime, i.e., $g_j<1$, we can make the approximations $C_{j}\approx 1$ and $S_{j}\approx g_{j}$ ($j\in\{1,2\}$). 
This leads to Eqs. (1a)--(1c) of the main paper.

\section{Fisher information}

Direct application of Eqs. (1) and (2) of the main text allows us to write
\begin{subequations}
\begin{align}
    FI_A &= \frac{4\eta_A T_A T_B g_1^2 g_2^2\sinphi^2}{T_A g_1^2 + g_2^2 + 2\sqrt{T_A T_B}g_1 g_2 \cosphi},\label{subeq:fi_a}\\
    FI_B &= \frac{4\eta_B T_A T_B g_1^2 g_2^2\sinphi^2}{T_B g_1^2 + g_2^2 + 2\sqrt{T_A T_B}g_1 g_2 \cosphi},\label{subeq:fi_b}\\
    FI_{CC} &= \frac{4\eta_A\eta_B T_A T_B g_1^2 g_2^2\sinphi^2}{T_A T_B g_1^2 + g_2^2 + 2\sqrt{T_A T_B}g_1 g_2 \cosphi}\label{subeq:fi_cc}.
\end{align}
\end{subequations}
Maximizing the FI over all possible phases $\phi$ yields Eqs. (6a)--(6c) of the main text.

\section{Modelling of the SU(2) interferometer}

We model the SU(2) interferometer similarly to the SU(1,1) in Section \ref{supp:sec:su11calc}. 
In particular, we replace the two PDC stages $PDC_{1,2}$ with two beam splitters. The first, input beam splitter has a reflectivity of 50\%, while the second, output beam splitter has a variable reflectivity $R$. In this case, the evolution of modes $\hat{a}$ and $\hat b$ is given by
\begin{subequations}
\begin{align}
    \hat{a}_\text{out} =& -\sqrt{1-\eta_A} \hat{d}_A 
    - \sqrt{\eta_A(1-T_A)(1-R)}  \hat{l}_A 
    - \sqrt{\frac{\eta_A}{2}} \bqty{\sqrt{(1-R)T_A} e^{i\phi} - \sqrt{R T_B}}\hat{a}\notag \\
    & + \sqrt{\frac{\eta_A}{2}} \bqty{\sqrt{(1-R)T_A} e^{i\phi} + \sqrt{RT_B}}\hat{b}
    -  \sqrt{\eta_A(1-T_B)R} \hat{l}_B^\dag,\\
    \hat{b}_\text{out} =&  
    \sqrt{\eta_B(1-T_A)(1-R)}  \hat{l}_A 
    - \sqrt{\frac{\eta_B}{2}} \bqty{\sqrt{RT_A} e^{i\phi} + \sqrt{(1-R)T_B}}\hat{a}\notag \\
    & - \sqrt{\frac{\eta_B}{2}} \bqty{\sqrt{R T_A} e^{i\phi} - \sqrt{(1-R)T_B}}\hat{b}
    -  \sqrt{\eta_B(1-T_B)(1-R)} \hat{l}_B^\dag
    -\sqrt{1-\eta_B} \hat{d}_A.
\end{align}
\end{subequations}

Considering the input state $\ket{\alpha, 0}$ for modes $A$ and $B$, the mean photon number (which coincides to the click probability for $\abs{\alpha}^2\ll 1$) takes the form
\begin{align}
    p_A &= \expval{n_A} = \frac{\eta_A}{2}\abs{\alpha}^2\bqty{\pqty{1-R}T_A + RT_B-2\sqrt{\pqty{1-R}RT_AT_B}\cos\pqty{\phi}},\\
    p_B &= \expval{n_B} = \frac{\eta_B}{2}\abs{\alpha}^2\bqty{\pqty{1-R}T_B + RT_A-2\sqrt{\pqty{1-R}RT_AT_B}\cos\pqty{\phi}}.
\end{align}
Therefore, the maximum attainable FI is given when the reflectivity $R$ balances the internal losses of the interferometer, i.e., when
\begin{equation}
    R = \frac{T_{A/B}}{T_A+T_B}
\end{equation}
holds true.
This further implies
\begin{align}
    FI_\text{A, SU(2)} &= 2\abs{\alpha}^2\eta_A \frac{T_AT_B}{T_A+T_B},\\
    FI_\text{B, SU(2)} &= 2\abs{\alpha}^2\eta_B \frac{T_AT_B}{T_A+T_B},
\end{align}
meaning that the maximum FI in an SU(2) with a weak input coherent light is given by
\begin{equation}
    FI_\text{max, SU(2)} = 2\abs{\alpha}^2\eta_{\max} \frac{T_AT_B}{T_A+T_B}.
\end{equation}
where $\eta_{\max} = \max\{\eta_A,\eta_B\}$.

Now, if we assume that, for both SU(2) and SU(1,1), the sample is subjected to the same amount of photons, then $\abs{\alpha}^2/2 = g_1^2$ applies.
This means that 
\begin{equation}
    FI_\text{max, SU(2)} = 4g_1^2\eta_{\max} \frac{T_AT_B}{T_A+T_B}.
\end{equation}

For an SU(1,1) interferometer, we have seen in the main paper that the maximum efficiency achievable is the one of the best performing singles, i.e.,
\begin{equation}
    FI_\text{max, SU(1,1)} = 4g_1^2\eta_{\max}T_AT_B.
\end{equation}

Therefore, for equal resources impinging on the sample, low mean photon number and equal internal and external losses (\textit{conditional advantange}), the SU(1,1) interferometer has higher FI than an SU(2) interferometer with coherent input when
\begin{equation}
    FI_\text{max, lossy SU(1,1)} > FI_\text{max, lossy SU(2)} \iff T_A+T_B>1
\end{equation}
holds true.
On the other hand, when comparing a lossy SU(1,1) interferometer with a lossless SU(2) interferometer with identical detection efficiency (\textit{unconditional advantage}), the SU(1,1) interferometer outperforms the SU(2) interferometer when
\begin{equation}
    FI_\text{max, lossy SU(1,1)} > FI_\text{max, lossless SU(2)} \iff 2T_AT_B>1
\end{equation}
applies.

\end{document}